\documentclass[12pt]{article}
\pdfoutput=1
\usepackage{amssymb,bm}
\usepackage{color,graphicx}
\usepackage{amsmath}

\newcommand{\rmd}{\mathrm{d}} \newcommand{\rmi}{\mathrm{i}} \newcommand{\rme}{e}
\newcommand{\tb}{\mathring{\tau}}

\newcommand{\cb}{\mathring{c}}
\newcommand{\Or}{{O}}
\newcommand{\litor}{{\mathrm o}}

\newcommand{\fex}{f_{\mathrm{ex}}}
\newcommand{\fb}{f_{\mathrm{b}}}
\newcommand{\fs}{f_{\mathrm{s}}}
\newcommand{\fres}{f_{\mathrm{res}}}
\newcommand{\tc}{T_{\mathrm{c}}}
\newcommand{\sk}{\sf k}
\newcommand{\regsol}{\varphi} 

\newcommand{\aseq}{\simeq}
\newcommand{\asprop}{\sim}

\newcommand{\zL}{\mathfrak{z}}
\newcommand{\zm}{{\sf z}}
\begin{document}
\thispagestyle{empty}
\hyphenation{wave-length}
\title{The three-dimensional  $O(n)$ $\phi^4$ model on a strip with free boundary conditions: exact results for a nontrivial dimensional crossover in the limit $n\to\infty$\footnote{Invited talk at MQFT15}}
\author{H.~W.\ Diehl and Sergei B.\ Rutkevich\\
Fakult\"at f\"ur Physik, Universit\"at Duisburg-Essen,\\ D-47058 Duisburg, Germany}
\maketitle \thispagestyle{empty}
\begin{abstract}
We briefly review recent  results of exact calculations of critical Casimir forces of the $O(n)$ $\phi^4$ model as $n\to\infty$ on a three-dimensional strip bounded by two planar free surfaces at a distance $L$. This model has long-range order below the bulk  critical temperature $\tc$ of the bulk phase transition only in the limit $L\to\infty$ but remains disordered for all $T>0$ for an arbitrary finite strip width $L<\infty$. A proper description of the system scaling behavior near $\tc$ turns out to be  a quite challenging problem because in addtion to bulk, boundary scaling, and also finite-size critical behaviors, a nontrivial dimensional crossover must be handled. The  model admits an exact solution in the limit $n\to\infty$ in terms of the eigenvalues and eigenenergies of a self-consistent Schr\"odinger equation. This solution contains a potential $v(z)$ with the near-boundary singular behavior $v(z\to 0+)\approx -1/(4z^2)+4m/(\pi^{2}z)$, where $m=1/\xi_+(|t|)$ is the inverse bulk correlation length and $t\sim (T-\tc)/\tc$, and a corresponding singularity at the second boundary plane. In recent joint work with colleagues, the potential $v(z)$, the excess free energy, and the Casimir force were obtained numerically with high precision. We explain how these numerical results can be complemented by exact analytic ones for several quantities (series expansion coefficients of $v(z)$, the scattering data of $v(z)$ in the semi-infinite case $L=\infty$ for all  $m\gtreqless 0$, and the low-temperature asymptotic behavior of the residual free energy and the Casimir force) by a combination of boundary-operator and short-distance expansions, proper extensions of inverse scattering theory, new trace formulas, and semiclassical expansions.
\end{abstract}
DOI: 10.1134/S004057791702009X

\section{Introduction}

Neutral atoms such as ${}^4$He interact via long-range interactions --- the very familiar van-der-Waals interactions. Their origin is well understood (see, e.g., ~\cite{CL95}). Because of quantum mechanical charge fluctuations, either atom has a fluctuating dipole moment producing an electric field that the other atom experiences.  Although the averages of the dipole moments are zero, the values of their squares are nonzero, and a fluctuation-induced (van-der-Waals) interaction consequently results. In 1948, Casimir showed  that analogous fluctuation-induced interactions occur between  macroscopic metallic objects in empty space because of the disturbance of the electromagnetic ground-state energy  by these objects \cite{Cas48}. Considering a pair of parallel grounded metallic plates of surface area $A$ at a separation $L$ in vacuum, he found that the (unmeasurable, infinite) electromagnetic vacuum energy per surface area $A$ ($A\to \infty$) acquires an $L$-dependent contribution that does not contain ultraviolet (UV) divergences,
\begin{equation}
\Delta E_0/A=\frac{\hbar c}{2}\left[\int\frac{\rmd^2p}{(2\pi)^{2}}\sum_{k,\lambda}\sqrt{p^2+k^2}-2\int\frac{\rmd^3q}{(2\pi)^3}\,q\right] 
=-\frac{\pi^2}{720}\,\frac{\hbar c}{L}\,\frac{1}{L^2}.
\end{equation}
Here $(\bm{p},k)$ and $\bm{q}$ are three-dimensional wave vectors; $\bm{p}$ and $k$ are the two-dimensional and one-dimensional components respectively parallel  and  perpendicular to the plates. The boundary conditions that the electromagnetic fields must satisfy on the plates imply that $k$ is quantized, taking the values $k=\nu\pi/L$, $\nu=0,1,2\ldots$.  The sum $\sum_\lambda$ runs over one or two polarizations depending on whether $\nu=0$ and $\nu>0$.

The form of the result is easily understood. The factor $\hbar c/L$ sets an energy scale. Dimensional considerations therefore imply that the change of the ground-state energy per area have the form $\Delta E_0/A=\Delta_{\text{C}}\,\hbar c L^{-3}$, where $\Delta_{\text{C}}$ is a pure number, called the Casimir amplitude. This amplitude is independent of microscopic details, i.e., universal. The associated Casimir force per unit area becomes
\begin{equation}
\frac{\mathcal{F}_C(L)}{A}=-\frac{\partial E_0/A}{\partial
     L}=-\frac{\pi^2}{240}\,\frac{\hbar c}{L^4}=-\frac{0.013001255\ldots}{(L/\mu\text{m})^4}\,\frac{\text{dyn}}{\text{cm}^2}.
\end{equation}
This force depends on gross features such as the space dimension (here 3), the dispersion relation for electromagnetic waves, and the geometry. It is tiny unless the separation drops below microns. Because such small separations are realized in  micromechanical and microelectromechanical devices, such Casimir forces are of technological interest and should be taken into account \cite{BR01}. An extensive account of the theory and measurements of Casimir forces and a list of references can be found in \cite{BKMM09}.

Thirty years after Casimir's pioneering work \cite{Cas48}, it was shown in \cite{fDG78} that analogous effective forces can be generated by long-wave\-length, low-energy classical  thermal fluctuations of systems near critical points  between macroscopic objects immersed into them such as plates and container walls. The subject of such critical Casimir forces (CCFs) has attracted considerable attention during the last 25 years. Early attempts to detect such forces focused on their indirect confirmation through their effects on wetting layers \cite{GC99}-\cite{FYP05}. Subsequently, Bechinger's group accomplished direct measurements of CCFs between a wall and colloidal particles immersed into binary fluid mixtures near their consolute points \cite{HHGDB08} -\cite{Gam09}.

Just as the Casimir forces of QED (QEDCFs), CCFs exhibit universal properties. They depend on gross features of the medium that mediates them, namely, the universality class of the associated bulk critical behavior, on gross features of the immersed macroscopic objects or walls, which are usually encoded in the large-scale boundary conditions of the continuum field theory, and the geometry. Despite obvious analogies between QEDCFs and CCFs, there are a number of crucial differences between them. Although previously noted (see, e.g.,  \cite{DS11} - \cite{DGHHRS14}), these differences are not always fully appreciated. A first obvious difference is that CCFs originate from thermal, not quantum, fluctuations. Quantum fluctuations do not contribute to the leading large-distance asymptotic behaviors unless there are fluctuation-induced forces near quantum critical points \cite{Sac11}.

Another important difference is that the universal properties of QEDCFs can normally be modeled by an effective free field theory in which the interaction of the electromagnetic fields with  macroscopic objects such as metallic plates are taken into account via boundary conditions. In the case of CCFs, interacting classical field theories such as $\phi^4$ theories must usually be considered. This is dictated by the requirement that the bulk critical behavior is properly described. 

One more difference is that unlike QEDCFs, CCFs are not normally fully fluctuation induced, because the order parameter $\phi$ can acquire a nonzero average in a strip geometry, either in the low-temperature regime, if the system has long-range order (LRO), or even generically at all temperatures as in the case of classical fluids and binary fluid mixtures because of an explicit breaking of the $\phi\to-\phi$ symmetry by boundary fields \cite{Die86a}, \cite{Die97}. If $\langle\phi\rangle\ne 0$, then this mean order-parameter profile reacts to changes in the separation between boundary planes, thus producing an effective force even in the Landau theory. If so, the CCF consists of a non-fluctuating background plus fluctuation-induced contributions.

Here, we focus on the case of CCFs in systems whose critical behavior can be described by an $O(n)$ invariant Hamiltonian. Our interest in this problem derives from several sources. First, the $O(2)$ case directly relevant for experiments that measure CCFs in ${}^4$He near the superfluid $\lambda$ transition via the thinning of wetting layers \cite{GC99}, \cite{GSGC06}. Second, we believe that this problem represents a much closer analog to QEDCFS than that of binary fluid mixtures. This is because the order-parameter profile $\langle \phi(\bm{r})\rangle$ of ${}^4$He  remains zero for all temperatures $T>0$ if the thickness $L$ of the strip is finite. Third, we believe this challenging problem is fundamentally important. What makes it quite hard is that it involves difficulties beyond those usually  encountered in studies of CCFs in systems with free boundaries, namely, the requirement to deal simultaneously with bulk, boundary, and finite-size critical behaviors. An additional complication to cope with in the case of  three-dimensional strips with  $O(n)$ symmetric Hamiltonian is the subtle dimensional crossover of a three-dimensional bulk ($L=\infty$) system with LRO at low temperatures to an $L<\infty$ strip with no LRO at any $T>0$ \cite{DGHHRS12}, \cite{DGHHRS14}, \cite{RD15a}.

Our purpose here is to elucidate these problems and show how they can be handled with the same approach, namely, the exact solution for $n\to\infty$. We explain why the $n=\infty$ solutions for the scaling functions of the residual free energy and the CCF under free boundary conditions (FBCs) are considerably harder to obtain than their analogues under periodic boundary conditions (PBCs) and that the former qualitatively and quantitatively behave differently from the latter.

The remainder of this paper is organized as follows. In the next section, we introduce the $O(n)$ $\phi^4$ model on a strip, specify the boundary conditions, and discuss its finite-size phase diagram. We then recall that the model can be solved exactly in the  $n\to\infty$ limit in terms of the eigenvalues and eigenstates of a self-consistent one-dimensional Schr\"odinger equation.

\section{Model, boundary conditions, and background}

\subsection{Hamiltonian and boundary conditions}

We consider the model defined by the reduced Hamiltonian
\begin{equation}\label{eq:Ham}
\mathcal{H}[\bm{\phi}]=\int_{\mathfrak{V}}\!\rmd^dr\,\Big[\frac{1}{2}(\nabla\bm{\phi})^2+\frac{\tb}{2}\phi^2+\frac{g}{4! n}\phi^4\Big]+\frac{\cb}{2}\int_{\mathfrak{B}_1\cup\mathfrak{B}_2}\!\rmd^{d-1}y\,\phi^2\;,
\end{equation}
where $\bm{\phi}(\bm{r})=(\phi_\alpha(\bm{r}))$, $\alpha=1,\dotsc,n$,  is a real-valued $n$-component order-parameter field defined on the $d$-dimensional strip $\mathfrak{V}=\mathbb{R}^{d-1}\times[0,L]$. We write the position vector  as $\bm{r}=(\bm{y},z)$ where  $\bm{y}\in\mathbb{R}^{d-1}$ and $z\in[0,L]$ are coordinates respectively parallel and perpendicular to the boundary planes $\mathfrak{B}_1=\{\bm{y},0)\}$ and $\mathfrak{B}_2=\{(\bm{y},L)\}$.  Here $(\nabla\bm{\phi})^2$ briefly denotes $\sum_\alpha(\nabla\phi_\alpha)^2$ and $\phi^2$ denotes the square $\sum_\alpha\phi_\alpha^2$ of the length of $\bm{\phi}$. 

The boundary terms of the Hamiltonian 
imply the so-called Robin boundary conditions \cite{Die86a}, \cite{Die97}, \cite{SD08}
\begin{equation}\label{eq:Robinbc}
\partial_n\bm{\phi}=\cb\,\bm{\phi},\;\;\bm{r}\in\mathfrak{B}_1\cup\mathfrak{B}_2,
\end{equation}
where $\partial_n$ means the derivative along the inner normals. For simplicity, we choose the same value of $\cb$ on both boundary planes. Relaxing this condition by choosing different values of $\cb_j$ on $\mathfrak{B}_j$ would be straightforward \cite{SD08,DS11,DGHHRS14,RD15a}  but is unnecessary here because  we focus on the case where the large-distance behavior is governed by a fixed point corresponding to Dirichlet boundary conditions on both planes $\mathfrak{B}_1$ and $\mathfrak{B}_2$, which can be studied by setting $\cb=\infty$ \cite{Die86a}, \cite{Die97}, \cite{SD08}, \cite{DS11}.

The model with PBCs along the $z$-direction,
\begin{equation}\label{eq:PBC}
\bm{\phi}(\bm{y},L)=\phi(\bm{y},0),
\end{equation}
 is defined similarly: in this case the boundary term involving $\int_{\mathfrak{B}_1\cup\mathfrak{B}_2}$ in Eq.~\eqref{eq:Ham}  is absent because there are no boundaries. 
\subsection{Remarks on UV singularities and renormalization issues}

We consider these models for $2<d<4$. We mainly focus on the ($d=3$)-dimensional case. For $d<4$, these models are superrenormalizable. The UV singularities in multipoint cumulants can be eliminated by introducing the renormalized quantities 
\begin{equation}
\tau=\tb-\tb_c
\end{equation}
and (only for FBCs)
\begin{equation}
c=\cb -\cb_{\mathrm{sp}}.
\end{equation}
where the UV-diverging parameters $\tb_c$ and $\cb_{\mathrm{sp}}$ are defined by the location of the critical point.

If we use a large-momentum cutoff $\Lambda$ to regularize, then these shifts vary as $\tb_c\sim \Lambda^{d-2}$ and $\cb_{\text{sp}}\sim\Lambda^{d-3}$.
Additional (logarithmic) UV singularities appear at $d=4$, which require further reparametrizations. The interested reader can find the relevant details concerning this issue and how to renormalize the theory in $4-\epsilon$ dimensions in the review articles \cite{Die86a},  \cite{Die97} and in \cite{DD80},  \cite{DD81a},  \cite{DD81b},  \cite{DD83a},  \cite{SD08},  \cite{DS11}. Because we restrict ourself to the case $2<d<4$, we can disregard this matter. But the bulk and surface free energies  require subtractions (additive counterterms) to cancel their UV singularities  even when $d<4$. We restrict ourself to occasional remarks at appropriate places and in a few cases cite more detailed expositions in the literature. 

\subsection{Bulk, surface, excess, and residual free energies}

Keeping in mind that quantities such as free energies and the partition function require a regularization of UV singularities to be well defined, we introduce the partition functions formally through the functional integral
\begin{equation*}
\mathcal{Z}=\int\mathcal{D}[\bm{\phi}]\,\rme^{-\mathcal{H}[\bm{\phi}]}
\end{equation*}
and the reduced free energy $-A^{-1}n^{-1}\ln\mathcal{Z}$ per surface area $A=\int\rmd^{d-1}y\to\infty$ and number of components. We can then define the limit $n\to\infty$ of the free energy (area) density $f_L$, the bulk free energy density $\fb$, the excess free energy density $\fex$, the surface free energy $\fs$, and the residual free energy $\fres$ by
\begin{eqnarray}
f_L&=&-\lim_{n\to\infty} \lim_{A\to\infty}\frac{\log\mathcal{Z}}{nA}\;,\\\
\label{eq:fbdef}
\fb&=&\lim_{L\to\infty}\frac{f_L}{L},\\
\fex(L)&\equiv& f_L-L\,\fb,\\
\fs&=&\frac{\fex(\infty)}{2},\\
\fres(L)&=&\fex(L)-2f_{\mathrm{s}}.
\end{eqnarray}
The bulk free energy $\fb$ is independent of the boundary conditions (PBCs or FBCs). The  other four free energies differ for FBCs and PBCs. Specifically,  $\fs^{\text{FBC}}$, $\fex^{\text{FBC}}$, and $\fres^{\text{FBC}}(L)$ depend also on the surface interaction constant $\cb$, while $\fs^{\text{PBC}}\equiv 0$.

\subsection{Finite-size phase diagram for $d=3$}

\begin{figure}[htbp]
\begin{center}
\includegraphics[scale=1]{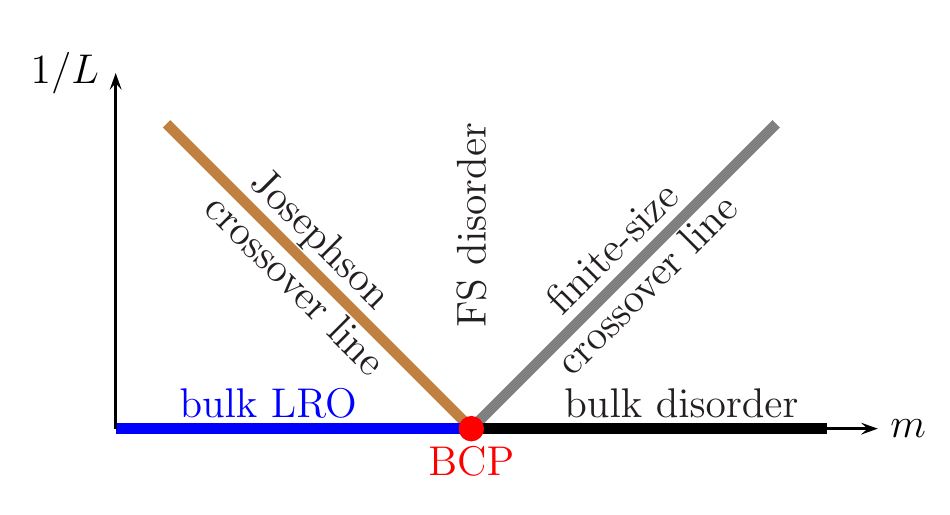}
\caption{Finite-size phase diagram for $d=3$: the bulk critical point (BCP) separates the low-temperature phase with bulk LRO from the disordered high-temperature bulk phase. The lines $L=\pm|m|$ are crossover lines. See the main text for further explanation.}
\label{fig:FSpd}
\end{center}
\end{figure}

Hamiltonian~\eqref{eq:Ham} of the $O(n)$ $\phi^4$ theory is invariant under the continuous $O(n)$ symmetry. According to the Mermin-Wagner theorem \cite{MW66}, such  a symmetry cannot be spontaneously broken at any temperature $T>0$ in bulk systems with short-range interactions if $d\le 2$.  But for $d>2$, a phase with long-range bulk order (bulk LRO) exists below a critical temperature $\tc>0$. Rigorous extensions of the Mermin-Wagner theorem imply that classical ferromagnetic $O(n)$ models on networks cannot have a spontaneous magnetization at any $T>0$ if random walks on the same structure are recurrent, i.e., return to their starting point with probability one \cite{MW94}. Because random walks in a three-dimensional rectangular cuboid of 
 cross-sectional area $A=\infty$ and finite thickness $L$ are recurrent, no LRO phase is possible for $d=3$ at any $T>0$ when $L<\infty$. At $d=3$, we therefore have the finite-size phase diagram displayed in Fig.~\ref{fig:FSpd} regardless of whether we choose  PBCs or FBCs.
 
 The LRO low-temperature and disordered high-temperature bulk phases are located on the line $1/L=0$. As the temperature variable, we introduce
 \begin{equation}
 m=\frac{\text{sgn}(t)}{\xi_{\mathrm{b}}(|t|)} \aseq \frac{\text{sgn}(t),|t|^\nu}{\xi_{\text{b}}^{(+)}},
 \end{equation}
 where $\xi_{\text{b}}^{(+)}$ is the amplitude of the bulk correlation length $\xi_{\mathrm{b}}\aseq \xi_{\text{b}}^{(+)}\,t^\nu$ in the disordered ($t>0$) phase and $\nu=(d-2)^{-1}|_{d=3}=1$ for $n=\infty$.
 
The lines $L^{-1}=\pm m$ are finite-size crossover lines: on scales $|m|^{-1}\gtrsim L$,  the system not longer exhibits three-dimensional bulk behavior but becomes effectively two-dimensional. In the FBC case, where the translation invariance along the $z$ direction is broken, the length $|m|^{-1}$ also characterizes the thickness of the boundary regions within which local densities deviate for $z\lesssim |m|^{-1}<<L/2$ and $L-z\lesssim |m|^{-1}\ll L$ from their bulk values, which they approach in the limits $z,L\to \infty$ with $z<L/2$, and $L-z,L\to \infty$  with $L-z<L/2$. The approach to bulk behavior is exponential in $L$ for $m>0$. Hence, in the FBC case, the deviations of local densities from their bulk values vary as $\sim\rme^{-2zm+\Or(\log z m)}$ as $z,L\to\infty$ with $0<m^{-1}\ll z\ll L/2$.  Similarly, excess quantities that vanish in the limit  $L\to\infty$ such as $\fex$, decay for FBCs as $\rme^{-2mL+\Or(\log m L)}$  at large $L$. The factor of $2$ in both exponential forms can be understood via perturbation theory: the disturbance caused by the presence of a free surface at $z=0$ contributes to the free two-point function that depends on the distance $2z$ of one of its points $\bm{r}=(\bm{y},z)$ from the image point $(\bm{y},-z)$. In the PBC case (no boundaries), we have $\fex$ $\sim\rme^{-2mL+\Or(\log m L)}$  instead. 

For $m<0$, there is LRO in the bulk when $d\ge 3$. Because of the spontaneous breaking of the $O(n)$ symmetry breaking, correlations decay algebraically rather than exponentially, and the usual moment-based definitions of a correlation length are hence inapplicable \cite{FBJ73}. An appropriate coherence length or helicity length $\xi_{\text{hel}}$  characterizing the asymptotic decay of the bulk correlation function $\langle \bm{\phi}(\bm{r})\cdot\bm{\phi}(\bm{0})\rangle$ can be defined in terms of the  familiar helicity modulus $\Upsilon(T)$ by $\xi_{\text{hel}}=[(k_{\text{B}}T/\Upsilon(T)]^{1/(d-2)}$. Because $\Upsilon\asprop |m|^{\nu (d-2)}$ as $m\to 0-$ for $2<d<4$, this helicity length (or Josephson coherence length) varies as $\xi_{\text{hel}} \asprop |m|^{-\nu}$ in this limit, where $\nu|_{n=\infty}=(d-2)^{-1}$. At $d=3$ specifically, the temperature variable $|m|$ corresponds to $1/\xi_{\text{hel}}$ (up to a normalization factor). 

The length $\xi_{\text{hel}} $ plays a role analogous to $\xi_+$ for $t>0$: When it becomes comparable to $L$, finite-size corrections become important, and in the FBC case, it characterizes the thickness of the boundary regions $z\lesssim \xi_{\text{hel}} \ll L$ and $L-z\lesssim \xi_{\text{hel}}  \ll L$.

\section{Exact $n\to\infty$ solution of the model }
\subsection{Self-consistent equations at $n=\infty$}
There are different ways to derive the equations that must be solved to obtain the exact $n\to\infty$ solution of our model (see, e.g., \cite{Vas98}, \cite{MZ03}). One possibility is to use the Hubbard-Stratonovich transformation
\begin{equation}
\rme^{-g\phi^4/4!n}\propto \int_{-\infty}^{\infty}\rmd \psi\rme^{\phi^2\,\rmi\psi/2-3n\,\psi^2/2g}
\end{equation}
to conclude that the model in the large-$n$ limit is equivalent to $n$ copies of a constrained Gaussian model  with a one-component field $\Phi(\bm{r})$ (representing any one of the components $\phi_\alpha$), where boundary conditions~\eqref{eq:Robinbc} or \eqref{eq:PBC} must be taken into account. This Gaussian model has the effective Hamiltonian
\begin{equation}
\mathcal{H}_{\text{eff}}[\Phi,\psi]=\frac{1}{2}\int_{\mathfrak{V}}\rmd^{d}r\Big[\Phi\left(-\nabla^{2}+\tb+
\rmi\psi\right)\Phi+\frac{3n}{g}\psi^{2}\Big],
\end{equation}
where $\psi(\bm{r})$ is an auxiliary field. In the limit  $n\to\infty$, the integral over $\psi$ reduces to the  value of the integrand at the saddle point. Unlike the PBC case, this saddle point depends on $z$ because the translation invariance along the $z$ direction is broken. Writing it  as 
\begin{equation}
\rmi\psi(z)=v(z)-\tb,
\end{equation}
we find that the free energy $f_L$ can be expressed as
\begin{equation}
f_L=\frac{1}{2}(2\pi)^{-(d-1)}\int d^{d-1}p \sum\limits_\nu\log (p^2+\varepsilon_\nu)-\frac{3}{2g}\int_0^L\rmd{z}\,[\tb-v(z)]^2
\end{equation}
in terms of the eigenvalues $\varepsilon_\nu$ and eigenfunctions $\varphi_\nu(z)$, $\nu=1,2,\ldots,$ of the self-consistent Schr\"odinger equation defined by 
\begin{equation}\label{eq:SEeve}
[-\partial_z^2+v(z)]\varphi_\nu (z)=\varepsilon_\nu\, \varphi_\nu(z)
\end{equation}
and the self-consistency condition
\begin{equation}\label{eq:sccondv}
 \tb-v(z)=\raisebox{-0.3em}{\includegraphics[height=2em]{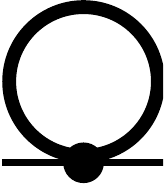}}=-\frac{g}{6}(2\pi)^{-(d-1)}\int d^{d-1}p\sum_\nu\frac{|\varphi_\nu(z)|^2}{\bm{p}^2+\varepsilon_\nu},
\end{equation}
where $\bm{p}\in\mathbb{R}^{d-1}$ is the wave vector conjugate to $\bm{y}$. 

Writing these equations, we implicitly assume the absence of LRO and hence $\langle\Phi(\bm{r})\rangle=0$. In the semi-infinite case, $L=\infty$, LRO would occur for $d>2$ whenever $\tb$ is below its bulk critical value $\tb_c=-(g/6)(2\pi)^{-d}\int d^{d}q \ q^{-2}$. But for $d>3$, the surface can be ordered for $\tb>\tb_c$ if $\cb$ exceeds a negative threshold value $\cb_{\mathrm{sp}}$. We do not consider the latter case, assuming $\cb$ to be sufficiently large for the order-parameter profile $\langle\Phi(\bm{r})\rangle$ to vanish for all $\tb$ or for all $\tb\ge\tb_c$ respectively depending on whether $L<\infty$ and $L=\infty$. 

\subsection{Scaling forms}
We are interested in the long-length-scale (scaling) solutions of Eq.~\eqref{eq:sccondv} both for PBCs and FBCs and want to use them to determine the scaling forms of the residual free energy $\fres$ and the Casimir force $\mathcal{F}_C$ as $L\to\infty$ and $m\to0$ with a fixed $x\equiv mL$. Because the potential $v(z)$ corresponds to the energy density, which scales as $m^{d-1/\nu}\mathop{=}m^2$ (for $n=\infty)$, we have
\begin{equation}\label{eq:vscalf}
v(z;L,m)=m^2\,v(|m|z;|m|L,\pm1)=\frac{v(z/L;1,mL)}{L^2}=\frac{v(1;L/z,mz)}{z^2}
\end{equation}
for FBCs. For PBCs, the $z$-dependence drops out, and only the first two equations are therefore applicable, with the first ($z$-related) arguments of the $v$ omitted. 

The residual free energy and the Casimir force take the scaling forms
\begin{eqnarray}
\fres(L,m)&\aseq&L^{-(d-1)}\,\Theta(mL)\,,\nonumber \\
 \mathcal{F}_C(L,m)&\aseq&L^{-d}\,\vartheta(mL),
\end{eqnarray}
with
\begin{equation}
\vartheta(x)=(d-1)\,\Theta(x)-x\,\Theta'(x),
\end{equation}
where $\Theta$ and $\vartheta$ are universal functions depending on $d$ and the boundary conditions.
\subsection{Exact $n=\infty$ results for periodic boundary conditions}
\begin{figure}[htbp]
\begin{center}
\includegraphics[scale=0.8]{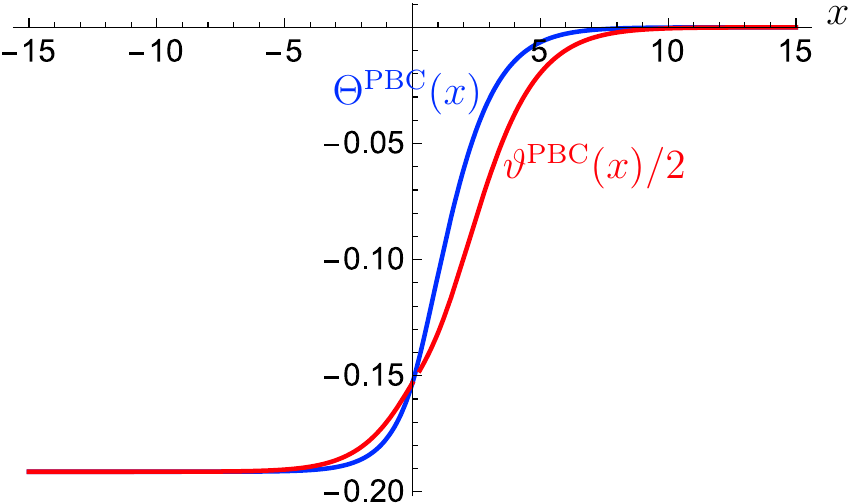}
\caption{Scaling function $\Theta(x)$ and $\vartheta(x)/2$ for $(d,n)=(3,\infty)$ and PBCs}
\label{fig:scfctprb}
\end{center}
\end{figure}

The $z$-independence of  the potential $v^{\text{PBC}}(L,m)$ for PBCs implies that the corresponding eigenvalues and eigenfunctions are exactly given by $\varepsilon^{\text{PBC}}_\nu=v^{\text{PBC}}+k_\nu^2$ and  $\varphi^{\text{PBC}}_\nu(z;L,m)=L^{-1/2}\,\rme^{\rmi k_\nu z}$ with $k_\nu=2\pi\nu/L$. At $d=3$ specifically, the scaled potential and the scaling functions can be determined in closed form \cite{Dan96}-\cite{DDG06}. We have
\begin{equation}
\check{v}(x)\equiv v^{\text{PBC}} (L, m) |_{\frac{{L=1,}}{{m=x}}} \mathop{=}^{d=3}\Big[2\,\mathrm{arcsinh}\frac{\rme^{x}}{2}\Big]^2
\end{equation}
and 
\begin{equation}
\Theta(x)\mathop{=}^{d=3}-\frac{8\pi^2x^3}{3}\,\theta(x)+Q\Big[\sqrt{\check{v}(x)}\Big]-\frac{\check{v}(x)}{12\pi}+\frac{x\,\check{v}(x)}{2}
\end{equation}
with
\begin{equation}
Q(y)\equiv\frac{1}{4\pi}\int_{y^2}^\infty dt \log \big[1-\rme^{-\sqrt{t}}\big]=-\frac{1}{2\pi}\big[\mathrm{Li}_3(\rme^{-y})+y\,\mathrm{Li}_2(\rme^{-y})\big],
\end{equation}
where $\theta(x)$ denotes the Heaviside function and $\mathrm{Li}_s(x)$ is the polylogarithm. The scaling functions $\Theta^{\text{PBC}}(x)$ and $\vartheta^{\text{PBC}}(x)/2$ for $d=3$ are shown in Fig.~\ref{fig:scfctprb}.
They are negative and monotonically decreasing for all $x$.  Their value 
\begin{equation}
\Delta^{\text{PBC}}_C\equiv \Theta_{d=3}^{\text{PBC}}(0)=\frac{1}{2}\,\vartheta_{d=3}^{\text{PBC}}(0)=-\frac{2\,\zeta(3)}{5\pi}= -0.1530506\ldots
\end{equation}
at $\tc$ defines the so-called Casimir amplitude $\Delta_C$. The $x\to-\infty$ saturation value 
 \begin{equation}
 \Delta^{\text{PBC}}_{G,C}=\Theta_{d=3}^{\text{PBC}}(-\infty)=\frac{1}{2}\,\vartheta_{d=3}^{\text{PBC}}(-\infty)=-\frac{\zeta(3)}{2\pi}=-0.191313298\ldots
 \end{equation}
  is the corresponding amplitude for a massless scalar free field theory.
  
  \subsection{General features of the scaling functions for free boundary conditions}
  
The behaviors of  $\Theta(x)\equiv\Theta^{\text{FBC}}(x)$ and $\vartheta(x)\equiv \vartheta^{\text{FBC}}(x)$ differ qualitatively and quantitatively from those of their PBC analogs. First evidence of this came from experimental investigations of the thinning of wetting layers of ${}^4$He in contact with copper substrates at and below the normal-to-superfluid $\lambda$ transition \cite{GC99}, \cite{GSGC06}. The scaling function $\vartheta_{{}^4\text{He}}(x)$ extracted from the wetting data turned out to be negative definite for all $x$, just like the $n=\infty$ result $\vartheta^{\text{PBC}}(x)$ shown in Fig.~\ref{fig:scfctprb}. But unlike the latter function, it did not monotonically decrease from $\vartheta(\infty)=0$ to its limit value $\vartheta(-\infty)$ as $x\to-\infty$ but passed through a minimum located at $x<0$ and subsequently seemed to increase monotonically to a limit value $\vartheta(-\infty)<0$ as $x$ decreased further. This behavior could be corroborated by Monte Carlo calculations for the $XY$ model on a ($d=3$)-dimensional film with FBCs along the short direction \cite{Huc07}-\cite{Has10a}. Although these experimental and simulation results apply to the $n=2$ case of ${}^4$He, the ${n=\infty}$ scaling function $\vartheta^{\text{FBC}}(x)$ is expected to exhibit the following qualitative features: (1) negative definiteness for all $x$, (2) a pronounced smooth minimum at a value $x_{\text{min}}<0$, and a (3) monotonic approach to a  limit value $\vartheta(-\infty)<0$. On the other hand, the ${n=2}$ (${}^4$He) case is special in that a Kosterlitz-Thouless transition occurs to the left of the minimum. It has been identified  by Monte Carlo simulations \cite{Has10a}, \cite{Has09a} but leaves hardly any visible trace in $\vartheta^{\text{FBC}}_{{n=2}}(x)$.

It is a major challenge to determine the scaling functions $\vartheta^{\text{FBC}}(x)$ and $\Theta^{\text{FBC}}(x)$ for all $x$ within a single approximation scheme such that all of their important properties listed above are obtained at least in a qualitatively correct fashion. Perturbative renormalization-group (RG) approaches based on the $\epsilon$-expansion about the upper critical dimension $d^*=4$ were found to give acceptable results for the disordered phase $x>0$  \cite{DS11}, \cite{SD08}, \cite{KD91}, \cite{KD92a} but normally cannot describe dimensional crossovers properly. Hence, the $\epsilon$-expansion fails for $L<\infty$ near a critical or pseudo-critical point of the film \cite{DS11}, \cite{DGS06}-\cite{DG09}. In the PBC case, the Landau theory erroneously predicts a critical point of the film right at $\tc$. Hence, the $\epsilon$-expansion fails at $\tc$ \cite{DGS06}. For FBCs, the film becomes critical in the Landau theory at an $L$- and $\cb$-dependent temperature $T_{c,L}<\tc$. We can set $\cb=\infty$ and  choose Dirichlet boundary conditions. The film critical point is then located at $\tb-\tb_c=-(\pi/L)^2$, where the eigenvalue $\varepsilon_1$ of the lowest mode vanishes.

\begin{figure}[htbp]
\begin{center}
\includegraphics[scale=0.8]{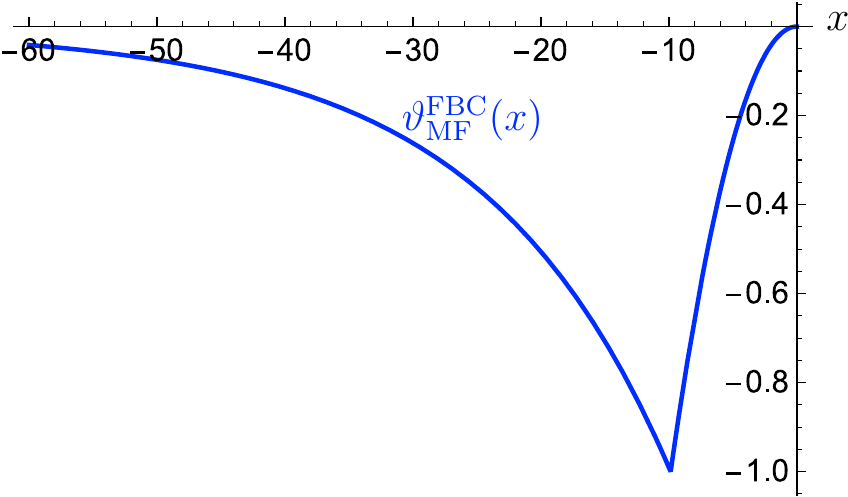}
\caption{Mean-field scaling function $\vartheta_{\text{MF}}^{\text{FBC}}(x)$ for FBCs: following \cite{MGD07}, we normalize its amplitude such that the minimum value at $x_{\text{min}}=-\pi^2$ is $-1$. This function vanishes for $x>0$, .}
\label{fig:MFvartheta}
\end{center}
\end{figure}

The mean-field (Landau theory) scaling function $\vartheta_{\text{mf}}^{\text{FBC}}(x)$ was determined exactly in \cite{MGD07}, \cite{ZSRKC07} and is plotted in Fig.~\ref{fig:MFvartheta}.

The critical temperature $T_{c,L}$ of the film translates into the location of the minimum of $\vartheta_{\text{MF}}^{\text{FBC}}(x)$. Because the ($\epsilon=4-d$)-expansion is based on an expansion about the Landau theory, it is clear that even if a critical point exists at  $T>0$ for $L<\infty$ (as it does in the $n=1$ case), close to that point, there must be a crossover to the effective ($d-1$)-dimensional long-distance behavior. Describing that crossover by the $\epsilon$-expansion method runs into serious difficulties. For $n\ge 2$, the situation is even worse because LRO is ruled out at $d\le 3$ for all $T>0$ when $L<\infty$. The Landau theory predictions are qualitatively wrong not only because it predicts a sharp transition at a $T_{c,L}>0$ for $L<\infty$ when $n\ge 2$ and $d\le 3$ but also because its predictions conflict with other important qualitative properties of $\vartheta_{{}^4\text{He}}^{\text{FBC}}(x)$ mentioned above. Unlike the latter function, $\vartheta_{\text{mf}}^{\text{FBC}}(x)$ is not continuously differentiable near the minimum: its derivative has a jump discontinuity at $x_{\text{min}}$. Furthermore, $\vartheta_{\text{mf}}^{\text{FBC}}(x)$ vanishes as $x\to-\infty$ instead of approaching a negative value. This is because the Landau theory ignores fluctuation effects due to Goldstone modes on scales $|m|^{-1}\lesssim L$. For the same reason, the Landau theory fails to predict the absence of LRO  at $T>0$ for $L<\infty$ in the continuous symmetry case $n\ge 2$ whith $d\le 2$. By contrast, the $n\to\infty$ limit complies with the Mermin-Wagner theorem \cite{MW66}, \cite{MW94}, providing a mechanism for nonperturbative mass generation.

Unfortunately, it is difficult to obtain exact analytical information about the potential $v(z)\equiv v^{\text{FBC}}(z)$ and the scaling functions $\Theta^{\text{FBC}}(x)$ and $\vartheta^{\text{FBC}}(x)$ for general values of $m$ and $L$, even at $d=3$ because  $v(z)$ depends on $z$. Accordingly, the spectrum $\{\varepsilon_\nu\}$ and eigenfunctions of Eq.~\eqref{eq:SEeve} are not generally known but  must be determined from Eq.~\eqref{eq:SEeve}, which must be solved self-consistently with   Eq.~\eqref{eq:sccondv} to obtain $v(z)$. All this together is a rather hard problem.

The scaling solution $v(z)$ of these equations for finite $L$ is not even known in closed analytic form at $\tc$. To gain information about $v(z)$, distinct methods  can and have been used, namely, (1) their direct solution for $m=0$ and $L=\infty$ \cite{BM77a}, (2) boundary-operator and short-distance expansions \cite{DR14}, (3) numerical solutions \cite{DGHHRS12}, \cite{DGHHRS14}, \cite{DGHHRS15}, \cite{DBR14}, (4) inverse scattering theory \cite{RD15a}, (5) trace formulas, (6) mapping the model for large $-x$ to a nonlinear $\sigma$-model \cite{DGHHRS12}, \cite{DGHHRS14}, and (7) semiclassical expansions \cite{RD15a}. In what follows, we briefly review the exact results obtained by these methods, focusing on the case $d=3$ unless stated otherwise.

\section{Survey of exact $n=\infty$ results for free boundary conditions}

The exact scaling solutions for the semi-infinite critical case $(m,L)=(0,\infty)$ for $2<d<4$ were found in \cite{BM77a} . For this, the authors used the ansatz $v(z)=Az^{-2}$, solved Eq.~\eqref{eq:SEeve} on the half-line $z\ge 0$ in terms of Bessel functions under the boundary condition that the eigenfunctions vary $\asprop z^{(1+\sqrt{1+4A})/2}$ for $z\to0$, and determined the coefficient $A$ from self-consistency condition~\eqref{eq:sccondv}: 
\begin{eqnarray} \label{eq:BMsol}
v(z;\infty,0)= \begin{cases} \frac{(d-3)^2-1}{4z^2},& 2<d<4,\\
\frac{(5-d)^2-1}{4z^2}, & 3<d<4,
\end{cases}\Bigg\}\qquad \mathop{=}\limits_{d=3} \frac{-1}{4z^2}.
\end{eqnarray}
The additional solution found  for $3<d<4$ applies to the so-called special surface transition where the variable $\cb$ takes the abovementioned threshold value $\cb_{\text{sp}}$ \cite{Die86a}. For $d=3$, only the solution associated with the so-called ordinary surface transition remains. The results agree with the rightmost scaling form given in Eq.~\eqref{eq:vscalf} and show that we must consider potentials that are singular at the boundary. Because the near-boundary behaviors as $z\to 0+$ and $z\to L-$  for general values of $L$ and $m$ must agree with Eq.~\eqref{eq:BMsol}, $v(z)$ behaves singularly at both boundary planes. At $d=3$, we must have
\begin{equation}\label{eq:vsurfsing}
v(z;L,m)\aseq \frac{-1}{4}\begin{cases}z^{-2},& z\to 0+,\\(L-z)^{-2},&z\to L-.
\end{cases}
\end{equation}

Schr\"odinger operators with singular potentials are somewhat subtle because it is unclear if a self-adjoint extension exists and what boundary condition must be imposed on the wave functions (see, e.g.,\ \cite{RS75}). An early investigation can be found in \cite{Cas50}, \cite{LL58}. In \cite{LL58}, the radial Schr\"odinger equation a$R''+2/rR-v(r)R=\varepsilon R$ was discussed for the potential $-\gamma/r^2$.  The value $\gamma=1/4$, which according to Eq.~\eqref{eq:BMsol} holds for $d=3$, turns out to be a threshold: physically meaningful self-adjoint extensions exist for $\gamma\ge 1/4$ \cite{KLP06}, \cite{RD15b} but not for $\gamma <1/4$, where no acceptable self-adjoint energy operator bounded from below results \cite{FLS71}.

We define two new scaling functions $\mathcal{V}_+$ and $\mathcal{V}_-$, whose index $\pm$ is determined by the sign of the parameter $m\gtrless 0$, by the formula
\begin{equation}
v(z;L,m)=z^{-2}\,\mathcal{V}_\pm(z/L,|m|z).
\end{equation}
Information about the behavior of $\mathcal{V}_\pm(\zL,\zm)$ for $\zL\to 0$ can be deduced via the boundary operator expansion using the fact that the contribution from the component $T_{zz}$ of the stress tensor yields the leading small-$\zL$ correction \cite{DR14}. For $d=3$, we obtain
\begin{equation}
\mathcal{V}_\pm(\zL,\zm)\mathop{=}\limits_{\zeta\ll 1} A_\pm(\zm)+\zL^3\,B_\pm(\zm)+\ldots,
\end{equation}
where the functions $A_\pm(\zm)$ and $B_\pm(\zm)$ are known to behave as
\begin{eqnarray}\label{eq:A}
A_\pm(\zm)&\mathop{=}\limits_{\zm\ll 1}&-\frac{1}{4}\pm\frac{4\zm}{\pi^2}+\frac{56\,\zeta(3)\,\zm^2}{\pi^4}+\litor(\zm^2),\\ \label{eq:B}
B_\pm(\zm)&\mathop{=}\limits_{\zm\ll 1}&\frac{256\,\Delta_C}{\pi}+\Or(\zm).
\end{eqnarray}
The first term on the right-hand side of Eq.~\eqref{eq:A} is known from \cite{BM77a}, and the second was derived in \cite{DR14} by an appropriate extension of the direct solution. The third term was determined in \cite{RD15a}  using inverse scattering theory together with a trace formula \cite{RD15b}. Finally, the result $B_\pm(0)$ given in Eq.~\eqref{eq:B} was obtained in \cite{DR14} using the boundary operator and short-distance expansions together with  results obtained in \cite{Car90b} and \cite{MO95}. The exact analytical  value of the Casimir amplitude it involves is unknown, but the numerical work of \cite{DGHHRS12}, \cite{DGHHRS14} yielded the high-precision result
\begin{equation}
\Delta_C=-0.01077340685024782(1).
\end{equation}

To achieve such a precision, a good understanding and control of corrections to scaling are necessary. At $d=3$, anomalous corrections $\asprop L^{-1} \log L$ appear because the bulk correction-to-scaling exponent $\omega|_{n=\infty}=4-d$ is degenerate with the surface correction-to-scaling exponent $d-(d-1)=1$ associated with the boundary operator $T_{zz}$ \cite{DGHHRS12}, \cite{DGHHRS14}, \cite{Die97}, \cite{DDE83}. The corrections to scaling can be partly suppressed by taking the limit $g\to-\infty$, where the model reduces to the  spherical model  in \cite{DBR14} with separate constraints on the energy density for each layer $z$ \cite{DGHHRS15}. The remaining corrections to scaling can be largely absorbed by changing $L$ to an effective thickness $L_{\text{eff}}$. 

In \cite{DGHHRS12}, \cite{DGHHRS14}, Eqs.~\eqref{eq:SEeve} and \eqref{eq:sccondv} were solved numerically to obtain $v(z;L,m)$, the eigenvalues $\varepsilon_\nu$, and the eigenfunctions for different values of  $m\gtreqless 0$ and $L$, and then analyzed to determine the scaling functions $\Theta^{\/\text{FBC}}(x)$ and $\vartheta^{\/\text{FBC}}(x)$. The results exhibit all important qualitative features known from the ${}^4$He case. We do not reproduce these results here, but refer the interested reader to the original papers. Instead, we report some additional exact results and mention how these were obtained. 

One fruitful method is to use inverse scattering theory. We recall that inverse scattering theory aims at reconstructing the potential from scattering or spectral data. More specifically, we consider the semi-infinite case $L=\infty$. From $v(z;\infty,\pm|m|)$, we substract its bulk value $\theta(m)\,m^2$ to obtain potentials $v(z;\infty,\pm 1)-m^2\,\theta(m)$ that vanish as $z\to\infty$, and we then scale $m$ to $1$, introducing $\zm=z|m|$ and the rescaled wavenumber $\sk=k/|m|$. We can introduce two Jost solutions of the Sturm-Liouville equation implied by Eq.~\eqref{eq:SEeve} with the limit behavior $f(\zm,\pm \sk)\mathop{\aseq}\limits_{\zm\to\infty}\rme^{\pm\rmi \sk \zm}$ and then choose a linear combination, the ``regular solution,'' that satisfies the boundary condition $\regsol(\zm,\sk)=\sqrt{\zm}\,[1+\Or(\zm)]$. Its asymptotic behavior
\begin{equation}\label{eq:varphiassigma}
\regsol(\zm,\sk)\mathop{\aseq}_{\zm\to\infty}\frac{A(\sk)}{\sk}\sin[\sk\zm+\eta(\sk)],\quad A(\sk)=\rme^{\sigma(\sk)}
\end{equation}
defines the scattering amplitude $A(\sk)$ and  phase shift $\eta(\sk)$. In our case, these scattering data differ for $m=\pm 1$, which we indicate with the subscripts on $A_\pm(\sk)$ and $\eta_\pm(sk)$. 

\begin{figure}[htbp]
\begin{center}
\includegraphics[width=0.6\textwidth]{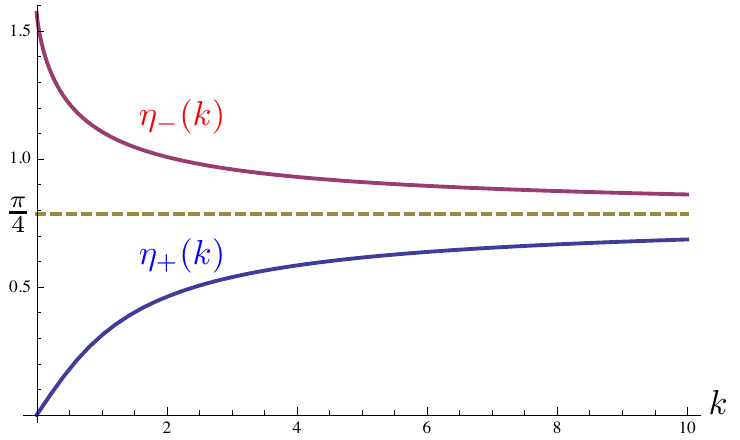}
\caption{\small Phase shifts $\eta_-(k)$ for $m=-1$ (red)  and $\eta_+(k)$ for 
$m=+1$ (blue), and $m=0$ (dashed yellow).}
\label{fig:eta}
\end{center}
\end{figure}

Two difficulties  had to be overcome to use inverse scattering theory. First, inverse scattering theory must be generalized to potentials with singular behaviors of the form $v(\zm)\aseq-\zm^{-2}/4+v_{-1}\zm^{-1}+\Or(1)$. Second, the scattering data from which to reconstruct the potentials are not given. We use the following trick. We introduce a UV-finite free-energy function $f[v]$ whose stationary equation $\delta f[v]/\delta v(z)=0$ yields Eqs.~\eqref{eq:sccondv} and \eqref{eq:SEeve}. We then consider variations $v(z)=v_*(z)+\delta v(z)$ about the stationary solution $v_*(z)$, which are chosen such that the singular part of $v(z)$ is unchanged. We then express
\begin{equation*}
\delta f[v_*,\delta v]=\int_0^\infty\rmd z\,\delta v(z)\,\delta f[v_*]/\delta v_*(z)
\end{equation*}
in terms of $\sigma_\pm(\sk)$ and $\delta\eta_\pm(\sk)$, taking into account that $\delta\sigma_\pm(\sk)$ and $\delta\eta_\pm(\sk)$ are related by Kramers-Kronig relations. This yields solvable integral equations for $\sigma_\pm(k)$, and we obtain
\begin{eqnarray*}
A_+(\sk)&=&\sqrt{\frac{\sk}{\text{arctan}\,\sk}},\quad \eta_+(\sk)=\int_0^\infty\text{d}u\,\frac{2\,\text{arctan}(\sk\tanh u)}{4u^2+\pi^2},\\
A_-(\sk)&=&\frac{|\sk|}{\sqrt{1+\pi|\sk|/2}},\\ \eta_-(\sk)&=&\text{sgn}(\sk)\bigg\{\frac{\pi}{2}+\frac{1}{2\pi}\bigg[\mathrm{Li}_2\Big(-\frac{\pi |\sk|}{2}\Big)-\mathrm{Li}_2\Big(\frac{\pi |\sk|}{2}\Big)\nonumber\\&&\strut -\log\bigg(\frac{\pi |\sk|}{2}\bigg)\log\frac{2-\pi |\sk|}{2+\pi |\sk|}\bigg]\bigg\}.
\end{eqnarray*}
The phase shifts $\eta_\pm(\sk)$ are plotted in Fig.~\ref{fig:eta}.

These results allow computing the scaling function of the surface two-point function  exactly. Because the order parameter varies as $\sqrt{z}$ for  $z\to 0$, an appropriate boundary operator that does not vanish at $z=0$ can be defined by $\Phi_s(\bm{y})\equiv \lim_{z\to 0}z^{-1/2}\,\Phi(\bm{y},z)$. For its correlation function, we obtain
\begin{equation}
\langle\Phi_s(\bm{y})\,\Phi_s(\bm{0})\rangle=\frac{1}{2\pi y^2}\left\{\theta(m)\,\rme^{-my}+\theta(-m)\left[1+|m|y+\frac{m^2y^2}{2}\right]\right\},
\end{equation}
and the bulk correlation function is given by
\begin{equation}
\langle\Phi(\bm{y},\infty)\,\Phi (\bm{0},\infty)\rangle=\frac{1}{4\pi y}\left\{\rme^{-my}\theta(m)+\theta(-m)\left[1+|m|y\right]\right\}.
\end{equation}

Building on the inverse scattering theory results, we can also compute the modified excess energy density for $m\le 0$  exactly:
\begin{equation}\label{eq:Jvarphi}
\int_0^\infty\rmd{z}\left[\langle\Phi(\bm{y},z)\rangle^2-\frac{m^3}{4\pi}\left(1-\frac{\theta(z|m|-1)}{2z|m|}\right)\right]=-\frac{1+\gamma_{\text{E}}+\log(4/\pi)}{2}\frac{m^2}{4\pi}.
\end{equation}
Here, the $1$ in the parentheses provides the usual subtraction of the bulk term $\langle\Phi(\bm{y},\infty)\rangle^2$. The additional subtraction associated with the term involving $\theta(z|m|-1)$ is necessary at $d=3$ because the energy density for $d>2$ has a power-law tail $\asprop z^{-(d-2)}|\mathop{=}\limits_{d=3}1/z$ due to spin waves \cite{Die86a}, \cite{RD15a}, \cite{DN86}.

Exact results can also be obtained for properties of the scaling functions $\Theta(x)$ and $\vartheta(x)$. The total free energy for $L<\infty$ is regular at $\tc$, but the residual free energy $\fres$ involves subtractions that are nonanalytic at $\tc$. Therefore, singular terms that compensate the mentioned nonanalytic terms must appear in the small-$x$ form of its scaling function $\Theta(x)$ t. These singularities have the form \cite{DR14}
\begin{equation}\label{eq:Thetasing}
\Theta(x)-\Delta_{\rm{C}}\mathop{\approx}_{x\to 0} -\left[\Delta A_0^{(\rm{s})}+\frac{x}{48\pi}\right]2x^2\,\theta(x)+\frac{1}{2\pi^3}\,x^2\log|x|+\ldots,
\end{equation}
where the ellipsis denotes regular terms of linear and higher orders in $x$ and $\Delta A^{(\mathrm{s})}_0$ is a universal amplitude difference associated with the thermal singularity of $\fs$. The latter is given by \cite{RD15a}
\begin{align}\label{eq:DelA0anexp}
\Delta A^{(\mathrm{s})}_0&=\int_0^\infty\frac{\rmd{u}}{16\pi}\,\frac{\coth(u)-1/u}{u^2+\pi^2/4}=\frac{1}{2\pi^3}\sum_{\sk=1}^\infty\frac{\log(2\sk)}{4\sk^2-1}
 =0.00944132199\ldots.
 \end{align}
 As an immediate consequence of Eq.~\eqref{eq:Thetasing}, we find that the second derivative of $\vartheta$ has the exact value
\begin{equation}
\vartheta''_3(0)=-\pi^{-3}.
\end{equation}

It is much harder to obtain the asymptotic behavior of $\Theta(x)$ for $x\to-\infty$, namely
\begin{equation}
\Theta(x)\mathop{\simeq}\limits_{x\to -\infty} -\frac{\zeta(3)}{16\pi}\Big[1+\frac{d_1 +2\log|x|}{|x|}+\litor(|x|^{-1})\Big],
\end{equation}
where
\begin{equation}
d_1=2\left[\gamma_{\text{E}}+\log\frac{4}{\pi}\right]-1-2\frac{\zeta'(3)}{\zeta(3)}=0.967205644660601\ldots
\end{equation}
The reason is that the asymptotic behaviors are needed both in the near-boundary and inner regions, respectively $z< |m|^{-1}< L/2$ and $|m|^{-1}<z<L/2$. This can be done by using distinct semiclassical expansion in each of them and matching the solutions in the intermediate region \cite{RD15a}. Such a  procedure turned out to be necessary, in particular, for determining the coefficient $d_1$. In contrast, the coefficient of the anomalous term $\asprop |x|^{-1}\log |x|$ can also be determined by mapping to a nonlinear $\sigma$~model \cite{DGHHRS14}. 

We can also determine the asymptotic  behavior of the scaled eigenvalues
$\varepsilon_\nu(1,mL)=L^2\,\varepsilon_\nu(L,m)$ in the limit $x\equiv mL\to-\infty$. We obtain \cite{RD15a}
\begin{equation}
\varepsilon_{\nu}(1,x)\mathop{=}_{x\to-\infty}\begin{cases}
\frac{\rme}{\pi}\,|x|\,\rme^{-|x|+\litor(1/|x|^0)},&\text{for }\nu=1,\\
\frac{1}{ \pi^2(\nu-1)^2}\left[1+\frac{2}{\pi(\nu-1)}\,\frac{\log(2|x|)}{|x|}+\litor \Big(\frac{1}{|x|}\Big)\right], &\text{for }\nu>1.\end{cases}
\end{equation}
The leading terms correspond to the spectrum of a free field massless field theory on a strip subject to Neumann boundary conditions \cite{DGHHRS12}, \cite{DN86}. The limit $\Theta(-\infty)=- \zeta(3)/16\pi$ is the associated Casimir amplitude.

\section{Concluding remarks}

The scaling functions $\Theta(x)$ and $\vartheta(x)$ of the residual free energy and the critical Casimir force in a three-dimensional strip differ qualitatively for PBCs and FBCs. Unlike the case of PBCs for which exact analytic closed-from expressions for these functions in the limit $n\to\infty$ are known for all values of $x$, such expressions are unknown for FBCs, where we must depend on numerical techniques to determine these functions for general values $x\in (-\infty,\infty)$. Nevertheless, a wealth of exact analytic properties of these functions 
have been derived using a combination of tools such as direct solution of the selfconsistent Schr\"odinger equation, boundary operator and short-distance expansions, inverse scattering theory methods, mapping to the nonlinear $\sigma$-model, trace formulas, and matched semiclassical expansions. These exact result have provided helpful benchmarks for assessing the precision of existing and potential future numerical results \cite{DGHHRS12}, \cite{DGHHRS14}, \cite{DGHHRS15}.

\section{Acknowledgments}
The authors are grateful to M.\ Hasenbusch, A.\ Hucht, and  F.\ Schmidt for fruitful interactions.

\end{document}